\documentclass[12pt,preprint]{aastex}

\shorttitle{Bar/update}
\shortauthors{L\'opez-Corredoira et al.}

\begin{document}


\title{The Long Bar in the Milky Way.\\ 
Corroboration of an old hypothesis}


\author{M. L\'opez-Corredoira$^{1}$, A. Cabrera-Lavers$^{1,2}$, T. J. Mahoney$^1$,
P. L. Hammersley$^{1,2}$, F. Garz\'on$^{1,3}$, C. Gonz\'alez-Fern\'andez$^1$}
\affil{$^1$ Instituto de Astrof\'\i sica de Canarias. C/.V\'\i a L\'actea, 
s/n, E-38200 La Laguna (S/C de Tenerife), Spain \\
$^2$ GRANTECAN, S. A., C/.V\'\i a L\'actea, s/n,
E-38200 La Laguna (S/C de Tenerife), Spain \\
$^3$ Departamento de Astrof\'\i sica, Universidad de La Laguna,
S/C de Tenerife, Spain}
\email{martinlc@iac.es}

%
%




\begin{abstract}
Recent GLIMPSE data have further confirmed 
the hypothesis of the existence of an in-plane
long bar different from the bulge of the Milky Way
with the same characteristics as emphasized
some years ago by our team.
In this paper, we present two new analyses that corroborate
recent and earlier claims concerning the existence in our Galaxy 
of a long flat bar with 
approximate dimensions 7.8 kpc $\times $1.2 kpc $\times $0.2 kpc 
and a position angle of approximately 43$^\circ $: 1) star counts
with {\itshape 2MASS} All-Sky Release and {\itshape MSX} data, 
which give an excess in the plane region along 
$0<l<30^\circ $ compared with $-30^\circ<l<0$, and which cannot be
due to the bulge, spiral
arms, a ring, or extinction; 2) new data on the distance of the long bar using the
red clump method, together with recent observations of our own
that are compared with our model, and that are in agreement
with the long-bar scenario.
\end{abstract}


\keywords{Galaxy: structure -- infrared: stars}


\section{Introduction}

Among the different stellar components of our Galaxy are 
a thin and thick disc, a halo, 
spiral arms, possibly a stellar ring and the bulge/bar.
The bar/bulge components (bulge, or bar, or bulge + bar) are perhaps 
the most difficult to disentangle from other components because of extinction and
our perspective of them due to our position in the Galactic plane  (Zhao 2000); they 
have been the topic of recent controversy concerning
their morphology (e.g.\ Sevenster et al.\ 1999;
Garz\'on 1999; Merrifield 2004, and references therein).
Our hypothesis in this debate has been that
there are indeed two components (Hammersley et al.\ 2001):
a bar which is long and in the plane ($-14^\circ <l<+30^\circ $,
$|b|<\approx 1.5^\circ $), and a triaxial bulge, which 
is short and much wider in latitude (observable at 
$|l|<\approx 15^\circ $, $|b|<\approx 10^\circ $).
Their position angles (angle between the major axis and the solar vector radius) 
might not be coincident, although 
the latest measurements give values that do not differ greatly: 43$^\circ \pm
7^\circ $ for the long bar (Hammersley et al.\ 2000; hereafter H00); and
28$^\circ \pm \sim 8^\circ $(syst.) for the triaxial bulge 
(L\'opez-Corredoira et al.\ 2005).

The story of the Galactic bar really begins with the discovery 
of large non-circular velocities  in the 3 kpc and  135 km/s spiral arms 
(Rougoor \& Oort 1960; Rougoor 1964). De Vaucouleurs (1964) was the first
to attempt to explain the motions of these arms in terms of a non-axisymmetric 
potential (a bar) at the centre of the Galaxy. He later predicted that the Milky 
Way was of type SAB(rs)bc on the grounds of the Galaxy's high spiral arm multiplicity,  
broken ring structure and non-circular H\ {\sc i} motions (de Vaucouleurs 1970). 
The non-axisymmetry hypothesis, or ``field model'', was not well received at the 
time by the astronomical community, which favoured Oort's (1977)  ``ejection model'', 
advocating the expulsion of gas from the centre of the Galaxy, even though it would 
require energies of the order 10$^{55}$ erg to power such an expansion of the spiral 
arms (van der Kruit 1971). Kerr (1967), however, explained the 3 kpc and 135 km/s 
arms in terms of a bar. Peters (1975) modelled the H~{\sc i} distribution 
derived from $l$--$v$ diagrams in the inner 4 kpc of the Galaxy in terms 
of concentric elliptical orbits. He 
found that an orientation of 45 deg w.r.t.\ to the solar radius vector fitted the 
maximum number of radio features in the $l$--$v$ diagrams, including the 3 kpc and 
135 km/s arms. With regard to molecular gas, Nakai (1992) compared CO ($J = 1\rightarrow 0$) 
emission maps of the barred spiral galaxies Maffei 2, NGC 2903 and NGC 253 with that of the 
Milky Way Galaxy (derived from the CO radial velocity at a sample of galactic longitudes). 
All four galaxies show a similar radial CO distribution with a sharp central peak and a 
secondary hump (corresponding to the bar--spiral arm transition). The radio evidence, then, 
gave strong indications of: a) a non-axisymmetric distribution of gas orbits and b) a CO 
intensity distribution that corresponded with other barred galaxies. 

Our hypothesis concerning the existence of a long bar originate from near-infrared Two Micron 
Galactic Survey data (TMGS, Garz\'on et al.\ 1993).
It began with an analysis of 
$K$-band TMGS star counts that led Hammersley et al.\ (1994) to posit the existence
of a bar of radius 4 kpc, with one of the tips at $l=27^\circ $,
and the other one at $l=-22^\circ $, which would give an angle of 
75 degrees \footnote{This was later retracted by
H00, who realized that the negative tip is in fact
at $l\approx -12^\circ $, so that the angle of the bar is instead $\approx 43^\circ $.}.
Calbet et al.\ (1995) presented ``tomographs'' of the Galactic plane
(derived from TMGS star counts and using the rather inaccurate assumption of a 
Dirac delta luminosity function, which could serve as a zeroth
order approximation of the structure of the Galaxy), and the long bar was
evident (figure 14 of Calbet et al.\ 1995), extending over a radius larger than
3--3.5 kpc, although the method was not accurate enough to provide 
information about its morphology. Calbet et al.\ (1996) also proposed
the existence of a dust lane preceding the bar at negative
longitudes, which would explain the higher extinction observed in this region.
Considerable progress was made by H00, who
measured the distance of several points along the bar, including $l=27^\circ $,
and, using a
technique of analysis of red-clump stars in near-infrared colour--magnitude
diagram, concluded that the angle was $43^\circ \pm 7^\circ $. 
The same analysis with the same data, only using 
near-infrared filters, was presented by Picaud et al.\ (2003), who
also compared the data with the Besan\c con model (Robin et al.\ 2003), 
and they obtained a similar result ($45^\circ \pm 9^\circ $).
Other papers of our team (Garz\'on et al.\ 1997; 
L\'opez-Corredoira et al.\ 1999, 2001b) have produced further evidence,
as will be described in Section \ref{.counts}. 

Concerning the stellar populations,
Ng (1998) finds two populations towards the Galactic Centre:
an old metal rich population ($Z=0.005$--0.08, $t$ = 13--15 Gyr), presumably
the old Galactic bulge, and a younger less metal-rich population
($Z=0.003$--0.03, $t=8$--9 Gyr), perhaps the long bar.
Cole \& Weinberg (2002) also showed a non-axisymmetric structure
that is likely to have been formed more 
recently than 3 Gyr ago and must be younger than 6 Gyr. However,
they are at $|b|>2^\circ $ so they were possibly  measuring the suburbs
of the long bar, some anomalously young population in the bulge,
or some contamination of the disc.

It is supposed that bars form as the result of an instability in differentially 
rotating discs (Sellwood 1981) whereas bulges are a primordial galactic 
component, so this distinction makes sense and it is not merely an
artefact invented to fit certain morphological features.
Athanassoula \& Misiriotis (2002), Athanassoula (2005)
and Athanassoula \& Beaton (2006)
use $N$-body simulations to produce long bars, which are broadened in 
at their centres into boxy bulges (the ratio of sizes between the semi-axis 
of the long bar and the major axis of the boxy bulge is approximately 1.5). 
Boxy bulges are in fact just a part of the long bar (Athanassoula 2005).
From a comparison of their $N$-body simulations and near-infrared photometry Athanassoula \& Beaton (2006)
argue that the boxy bulge of M31 extends into a long bar. 
There is a striking parallel with our own Galaxy, which
also has a boxy bulge (Picaud \& Robin 2004;
L\'opez-Corredoira et al.\ 2005) that extends
into the long bar described in this and previous papers.
Bars might be vertically extended (Kuijken 1996) but they might
be constrained to the plane too.
Indeed, many other galaxies exhibit both a bar and a bulge; e.g.\
NGC 3351, NGC 1433 and NGC 3992, whose images in 
the Palomar plates, for instance, show the existence of both
structures.

Nevertheless, many authors (reviewed by Merrifield 2004) 
have for many years ignored the bar + bulge hypothesis  
and preferred to speak about a unique non-axisymmetric
structure, namely a bar (i.e.\ bulge) that is triaxial, short and fatter
(what we call this component the bulge). These authors did 
not consider the possibility of a long bar constrained in the Galactic plane. Although
they were more or less successful in fitting the stars in off-plane regions,  
the abundance of new data now indicates
that some new component in the Galactic plane at longitudes
$+15^\circ <l<+30^\circ $ is being observed that is unexplainable
in terms of a disc + bulge + extinction model. 
This fact will be corroborated in \S \ref{.counts} with All-Sky {\itshape 2MASS} 
and {\itshape MSX} data.
In addition, in \S \ref{.redclump} we use red clump star
data from the recent literature (including 
new observations of our own: Cabrera-Lavers et al.\ 2006b) 
to reinforce the old picture of our
hypothesis that there is a long bar apart from the bulge. Furthermore,
the position angle obtained in the past is confirmed
by the latest data (Benjamin et al. 2005). 

\section{The asymmetry in the counts}
\label{.counts}

L\'opez-Corredoira et al.\ (2001b) used star counts and extinction
maps of negative latitudes and confirmed the general picture of a long bar: 
higher counts in the positive than in negative
longitudes (for $|l|<30^\circ $ and strictly in the plane $b<\approx 1.5^\circ$).
This north/south asymmetry was confirmed with recenter GLIMPSE
data by Benjamin et al.\ (2005), with 25\% more sources in the north,
although they do not include the region $|l|<10^\circ $.
L\'opez-Corredoira et al.\ (2001b) also show a higher extinction 
on average iat negative latitudes. Babusiaux
\& Gilmore (2005) say, however, that they see no asymmetry in the
extinction, but we think they have only four
points to measure the asymmetry, and one of them ($l\approx -10^\circ $) 
has a minimum according to L\'opez-Corredoira et al.\ (2001b, Fig. 8). 
When the star counts are statistically corrected for extinction by using the 
colours, a clean image of the negative latitudes of the Galaxy is obtained
(L\'opez-Corredoira et al. 2001b, fig. 10) that clearly shows the in-plane
bar extending up to $l\approx -13^\circ $, as expected by H00. 
The schematic representation in L\'opez-Corredoira et al.\ 
(2001b, fig. 3) summarizes the model.
Unavane \& Gilmore (1998) and Alard (2001) also explored 
asymmetries in star counts in the plane but only for $|l|<\approx 4^\circ $.

\begin{figure}
\begin{center}
\vspace{1cm}
{\par\centering \resizebox*{8cm}{6cm}{\includegraphics{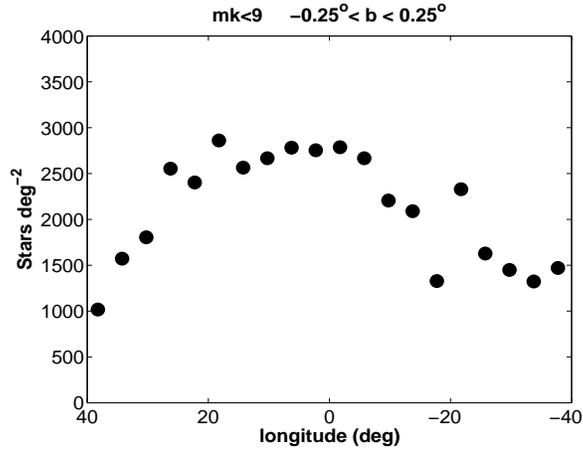}}\par}
\end{center}
\caption{2MASS star counts showing the asymmetry in the plane 
between positive and negative longitudes in $|l|<30^\circ $. 
Counts have been averaged over $\Delta l=5^\circ $}
\label{Fig:2mass}
\end{figure}

With the data of {\itshape 2MASS} All-Sky data release (Skrutskie et al.\ 2006) in the $K$ band,
we plot the star counts up to $m_K=9$ in Fig.\ \ref{Fig:2mass}, and 
show a clear asymmetry. We corroborate the results by L\'opez-Corredoira
et al.\ (2001b), who used DENIS and TMGS data for some regions (not all-sky)
to claim that a long bar must be present, and that this structure is beyond
the bulge region ($|l|<15^\circ $), extending to nearly $l=30^\circ $
in positive longitudes.

With {\itshape MSX} mid-infrared data (Egan et al.\ 1999)
we also see the same trend. Figure \ref{Fig:msx} clearly shows
the excess at positive galactic longitudes; it also
shows a deficit between $l=-15^\circ $ and $l=-18^\circ $, which indicates that the
bar at negative latitudes finishes at $l\approx -14^\circ $ (the feature at $l=-22^\circ $
is thought to be the 3 kpc ring; L\'opez-Corredoira et al.\ 2001b).

Garz\'on et al.\ (1997) and L\'opez-Corredoira et al.\ (1999) confirmed
spectroscopically that most of the brightest $K$-band stars at $l=27^\circ $
in the plane are supergiants, thereby reinforcing the argument that this region
contains a prominent star formation region possibly associated
with the end of the bar where it meets the Scutum spiral arm.

Freundenreich (1998) realized that there are significant residuals in 
the plane over $|l|<30^\circ $ after subtracting
the disc and the bulge in {\itshape COBE}/DIRBE data; however, he prefers
to attribute these residuals to patchy star formation in a ring
or spiral arms. 
Picaud (2004) also says that this star formation region might be associated
with some other structure such as a ring. However, as plotted in 
L\'opez-Corredoira et al.\ (2001b, fig.\ 6), a ring would produce a 
shape in the distribution of star counts different from that shown in Figure \ref{Fig:2mass}.
Moreover, the deficit of stars observed between $l=-15^\circ $ and $l=-18^\circ $
in Fig.\ \ref{Fig:msx} indicates the absence of any possible continuous structure
between $l=0$ and $l=-22^\circ $; a ring should be continous, whereas our bar
finishes at $l=-14^\circ $. The bar fits the star counts perfectly, whereas
the ring does not.
A spiral arm solution on the far side of the Galaxy (at distance 10--12 kpc)
or closer than the central region of the Galaxy (distances less than 5 kpc)
is also discarded because of the arguments
given by Hammersley et al.\ (1994) and because, as will be shown in \S \ref{.redclump},
the distance of this structure does not fit any ring or spiral arm possibility.

Extinction is also discarded by Hammersley et al.\ (1994) and L\'opez-Corredoira et
al.\ (2001b). Furthermore, a closer examination of extinction-corrected counts, 
for instance that in L\'opez-Corredoira (2005, 
fig.\ 1), shows a still  higher number of counts at positive longitudes.
In Fig. \ref{Fig:m0700b0} we show an example
from the data of L\'opez-Corredoira et al.\ (2005) with $m_e<7.0$. 
GLIMPSE (Benjamin et al.\ 2005) and {\itshape MSX}
mid-infrared data (Fig.\ \ref{Fig:msx}) also show the asymmetry and 
are much less affected by extinction.

\ \ 

\section{The red clump as a distance indicator}
\label{.redclump}

The red clump method is well known and explained in several papers
(e.g.\ Stanek et al.\ 1997; H00; L\'opez-Corredoira et al.\ 2002):
in a colour--magnitude diagram, the maximum of the density of the
red clump stars is identified with the major axis of the bar
along each line of sight.
This is not strictly true; it is a good approximation only
when the bar is thin, that is, a very elongated structure, a long
bar rather than a thick bulge. In our case, the bar is thin because
there is a dispersion of $\sigma \approx$ 0.5--0.6 mag for the red clump, 
including the intrinsic broadening of the red clump itself and the 
dispersion of distances (Fig.\ \ref{Fig:sigma}).
In a thick bulge, the maximum density
in the line of sight is at 
the tangential point of the line of sight with
the ellipsoids and may differ from the major axis position by 
$\sim$1 kpc, but the difference is only $\sim$50 pc for the
thin bar (appendix \ref{.dif}). Also, there is
a difference between the maximum density along the line of sight
and the maximum of the star counts of the red clump vs.\ the apparent
magnitude, which is again negligible for thin bars (25--50 pc) 
and somewhat more important for the thick bulge (100--300 pc)
(appendix \ref{.dif2}).
Some authors, e.g.\ Babusiaux \& Gilmore (2005), 
have not considered these effects and applied the 
method to fit the major axis of a thick-bulge
structure, which is very inaccurate.

By gathering the different data in the literature\footnote{Red 
clump distance data from Stanek et al.\ (1997) are not valid for 
our analysis because they are all more than 2 degrees away from the plane.} 
with measurements of the distance as function of galactic longitude in the Galactic
plane (constrained to $|b|<\approx 1^\circ $), we get 
Fig. \ref{Fig:redc}, which includes data from:

\begin{itemize}

\item H00: Datum at $b=0$, $l=27^\circ $ at a distance of 5.7 kpc.
Other data might perhaps be derived from their colour--magnitude
diagrams but, as explained in Cabrera-Lavers et al.\ (2006b),
the crowded regions in the plane limit the validity of the method.
 
\item Picaud et al.\ (2003). Data at $b=0$. We take the given
distances of 5.9 kpc at $l=26^\circ $ and $l=27^\circ $.

\item Nishiyama et al.\ (2005). These authors present many points between $l=-10^\circ $
and $l=+10^\circ $, and we take one for each degree for our plot.
Their data are for $b=+1^\circ $, but we assume they are at the same
distance as the $b=0$ data. They also use the
red clump method and find that the position angle for $|l|<4^\circ $
is different from the position angle derived with all the data $|l|<10^\circ $.
They attribute this to the existence of a new structure at $|l|<4^\circ $
but we believe that this could  possibly be the effect of a superposition of a bulge 
and a long bar. In Fig. \ref{Fig:redc}, we show how their data are roughly
compatible with a bar of position angle 40--45 degrees.

\item Babusiaux \& Gilmore (2005). Data at $b=0$ (except the point at
$l=0$, which is at $b=+1^\circ $). A value of $R_\odot =8$ kpc is adopted. 
Babusiaux \& Gilmore (2005) observed in the near infrared with the
CIRSI survey in the plane fields
at $l=0$, $\pm 5.7^\circ $, $\pm 9.7^\circ $. They calculate the distance
of the red clumps for the bar for each of the five points. Their results are that
all the points except $l=-9.7^\circ $ fit a bar with a position angle
of 22.5$^\circ $.
The field at $l=-9.7^\circ $ is chosen to contain the presence of a stellar
ring or pseudo-ring at the end of the bar. The field at $l=+5.7^\circ $
presents a wide dispersion of distance owing perhaps to the added presence 
of distant (i.e.\ at the other side of the bar) disc red clump stars.
We think that Babusiaux \& Gilmore are observing the long bar 
and that the angle that their data give is possibly compatible with
a position angle of around 40--45 degrees, as is shown in Fig.
\ref{Fig:redc}. However, they reject this hypothesis
on the grounds that, in their view, a bulge + bar would produce a spread in the
red clump distances; 
possible alternative explanations are that the bar might be predominant in the
plane, that the angle of bulge and bar might be very similar in the plane,
or that the spread real but low to separate both populations.

\item Benjamin et al.\ (2005). We calculate from their fig.\ 4, based
on GLIMPSE data, the equivalent distances, 
assuming $R_\odot =8$ kpc instead of 8.5 kpc 
(which gives $M_{[4.5]}=-2.02$ instead of $-$2.15 for the red clump giants),
and assuming in this filter 0.05 mag/kpc of extinction.
Between $l=+12^\circ $ and $l=+28^\circ $ every four degrees, except
$l=24^\circ $, which is in a high extinction region and no hump associated
with the red clump is observed. 
These GLIMPSE data have recently been analysed by Benjamin et al.\ (2005), 
who found that,
when plotting the power law exponent of the star counts in the mid infrared
(their Fig. 2),
a different structure distinct from the disc can be
observed for $|l|<\approx 30^\circ $, a hump superimposed on the
disc counts. They suggest this to be due to the red clump giants
of the long bar. They measured the position angle to be $44\pm 10^\circ $.
Indeed, fig.\ 1 of Benjamin et al.\ (2005) also shows quite clearly
how the star counts in the mid infrared (with very low extinction)
trace an in-plane structure between $l=-14^\circ $ and $l=+30^\circ $.
This rediscovery of the long-bar has revived interest in the
hypothesis.

\item Cabrera-Lavers et al.\ (2006b). We add new data for
$b$ between -1$^\circ $ and +1$^\circ $, and $l$ between +18$^\circ $ 
and +28$^\circ $. The source of the data is the same that
of H00 and Picaud et al.\ (2003) but with more recent observations:
TCS-CAIN data (Cabrera-Lavers et al.\ 2006a).
 
\end{itemize}

Figure \ref{Fig:redc} shows how the data fit the H00
bar with a position angle of $43^\circ \pm 7^\circ $.
The error bars of the points represented  are roughly 700 pc at 5.7 kpc
($l=27^\circ $)  from the Sun (H00) and somewhat higher
for lower values of $l$; many authors give lower errors for 
their points but we think they were over-optimistic in their estimation 
of systematic and random errors. Most of the points are within the
expected prediction for the H00 long bar.
Perhaps the dispersion of points in the data of Cabrera-Lavers et al.\ (2006b) 
indicates that the error in distance is higher than 700 pc.
There is one point (at $l=+9.7^\circ $: $x=887$ pc,
$y=5247$ pc) in Babusiaux \& Gilmore (2005) that does not fit
the long bar. This point is possibly in error or  contaminated by
the bulge. The presence of this point is
used by Babusiaux \& Gilmore (2005) to claim that the bar position angle is
$22.5^\circ $ (a fit made with only 3 points and rejecting the point
at $l=-9.7^\circ $), but in view of all the other
 data, it is more likely that we have to reject instead the 
point at $l=+9.7^\circ $. Values for the position angle lower than
35 degrees  fail to fit all these data.

This angle of around 40--45 degrees also agrees with previous estimates.
Peters (1975) modelled the H~{\sc i} distribution with an orientation of 45 deg. 
The first direct evidence of a large stellar bar was produced by Weinberg (1992), 
who used AGB stars from the {\itshape IRAS} Point Source Catalog (1985) to trace the stellar density 
distribution within the solar circle. In spite of his simplistic assumptions of a 
homogeneous AGB sample with a low dispersion in luminosity and dust extinction proportional 
to distance within the solar circle, Weinberg gave estimates of the bar orientation 
(36 $\pm$ 10 deg) and semilength ($\sim$5 kpc).
Sevenster et al.\ (1999) measured a position angle of
45 degrees when analysing the structure
in the plane (and 25 degrees in off-plane regions, which denotes
the absence of a bar and  only the presence of a triaxial bulge  away from 
the plane). van Loon et al.\ (2003) detected the long bar at a Galactocentric
radius of $R\ge 1$ kpc and measured a position angle, $\phi \sim 40^\circ $. 
Groenewegen \& Blommaert\ (2005) measured $\phi = 47\pm 17^\circ $
in the relationship between the zero point of the $K$-band period--luminosity 
relation of Mira variables and galactic longitude although they
included data between $b=-1.2^\circ $ and $b=-5.8^\circ$ 
within $|l|<12^\circ $, so this corresponds to the long bar + thick bulge.

\section{Other parameters of the bar}
\label{.otherp}

In Table~\ref{Tab:param}
we summarize the parameters of the bar.
The major axis semilength, 3.9 kpc, is derived assuming that the end
of the bar is at $l=27^\circ $, that the angle of the bar is 43$^\circ $, 
and that the distance to the Galactic Center is $R_\odot=8$ kpc. 
The thickness may be
derived, as we have  said, from the dispersion in the red clump stars, which
is roughly 0.5--0.6 mag (H00; Benjamin et al.\ 2005; Cabrera-Lavers 
et al. 2006b). In particular, a fit of the data with $|b|\le 1^\circ $
from Cabrera et al.\ (2006b) (plotted in Fig. \ref{Fig:sigma}) gives:

\begin{equation}
\sigma _{\rm red\ clump}=(0.640\pm 0.037)-(0.0044\pm 0.018)\ l(deg.)\ {\rm mag}
\end{equation}

We assume an intrinsic Gaussian dispersion of the red clump of $0.3\pm 0.1$ 
mag (L\'opez-Corredoira et al.\ 2002) and consider  the variation
in height along the line of sight of the bar within $|b|\le 1^\circ $ to be 
negligible. We also take into account that, because  the orientation
of the bar, the angle between the line of sight and the direction
perpendicular to the bar is $\alpha =90^\circ -43^\circ -l$ (so we must 
multiply the thickness along the line of sight by a factor $\cos \alpha $)
and the distance of the bar is $r(l)=
R_\odot \frac{\sin 43^\circ}{\sin (43^\circ +l)}$, so that

\begin{equation}
\sigma _{\rm thickness\ bar}\approx (1420\pm 170)-(12.6\pm 5.6)\ l(deg.)\ {\rm pc}
.\end{equation}
This would be the horizontal thickness of the bar 
(within which $\approx 68$\% of 
the stars are contained). For low values $l$ it becomes thicker because
it is mixed with the bulge. For galactic longitudes around 
$l\approx 20^\circ $, the thickness is $1170\pm 200$ pc.
Between $l=20^\circ $ and $l=30^\circ $ it is nearly constant.

The vertical thickness is derived when we examine the decrease
in the red clump stars in Cabrera-Lavers et al.'s (2006b) data, giving
a number of around 200 pc (i.e.\ a scaleheight of around 100 pc
both in the north and in the south). 

The mean density is derived by means of
\[
{\rm Excess\ counts}(m_K<9,l)\approx \omega [\Delta r(l)]r(l)^2
\phi [M_K<9-5\log _{10}r(l)+5-A_K(r(l),l)] \overline {\rho }
.\]
From Fig.\ \ref{Fig:2mass}, 
the excess counts in positive longitudes over negative
longitudes is $\approx 1000$ deg$^{-2}$ [surface: $\omega =1$ deg$^{-2}=
3.0\times 10^{-4}$ rad$^{-2}$], roughly independent of $l$.
$\Delta r(l)=\frac{\sigma _{\rm thickness\ bar}}{\sin (43^\circ +l)}$.
The extinction $A_K(r(l),l)\approx 1.2$ (L\'opez-Corredoira et al.\ 2001b),
roughly independent of $l$. For $l=20^\circ $, $r(l)=6.1$ kpc,
which gives $M_{K,\rm max}=9-5\log _{10}r(l)+5-A_K(r(l),l)=-6.1$.
Hence, 
\begin{equation}
\phi [M_K<-6.1] \overline {\rho}=7\times 10^{-5}\ {\rm star\ pc}^{-3}
,\end{equation}
 a factor 4 lower value than the rough calculations in L\'opez-Corredoira
et al.\ (2001b, \S 6.2), mainly due to the different thickness used
here.

Roughly speaking, if we assume
as a model for the bar a box of dimensions $7.8\times 1.17\times 0.3=2.7$
kpc$^3$ with average density of stars up to $M_K=-6.1$
equal to $\phi [M_K<-6.1] \overline {\rho}(l=20^\circ )=7\times 10^{-5}$ 
pc$^{-3}$, the total number of stars up to absolute
magnitude $-$6.1 is $\sim$$200000$.
For comparison, the bulge up to this absolute magnitude has
around 600\,000 stars (L\'opez-Corredoira et al.\ 2000, \S 6.5), so the bar has
a third as many stars as the bulge at this range of magnitudes.
These bright stars would represent a fraction $2\times 10^{-5}$ 
[from the integration of the bulge luminosity function in L\'opez-Corredoira
et al.\ 2000] of the total number of stars; that is,
if we assume a similar ratio for the bar, $\overline {\rho} \approx 
3.5$ star pc$^{-3}$.

\begin{table}
\caption{Approximate parameters of the long bar}
\begin{center}
\begin{tabular}{cc}
\label{Tab:param}
Parameter & Value  \\ \hline
Angle Galactic Center--major axis: & 43$^\circ $ \\
Major semiaxis length: &  3900 pc \\
Horizontal thickness at $l=20^\circ $: & 1170 pc \\
Vertical thickness at $l=20^\circ $: & 200 pc\\
Star density at $l=20^\circ $, $b=0$ with $M_K<-6.1$: & $7\times 10^{-5}$ 
pc$^{-3}$ \\
Total mass (assuming bulge stellar pop.): & $6\times 10^9$ M$_\odot $ \\ 
\hline
\end{tabular}
\end{center}
\end{table}

We cannot derive directly the mass of the bar from the present numbers.
We would need a stellar library with an exact knowledge of the population
of the long bar, which is a task for a future work. If we
assumed roughly that the populations of the thick bulge and
the long bar were similar, since the long bar has a third as many
bright stars as the thick bulge and the bulge has 
$1.7\times 10^{10}$ M$_\odot$ (Sevenster et al. 1999), the bar
would have $\sim 6\times 10^9$ M$_\odot$.
 
Since this article is concerned solely with star counts based on
NIR photometry, the kinematics of the bar cannot be discussed here, but other authors 
have studied it; e.g. Debattista et al.\ (2002), although these authors do
not distinguish between a thick bulge and a long bar. It
remains to be  studied whether or the thick bulge and the long bar have
the same pattern speed.
A project dedicated to measuring with accuracy the velocities of
the plane stars at $0^\circ <|l|<30^\circ $ would be warranted.
Forthcoming surveys such as {\it RAVE} (Steinmetz et al. 2006) 
will in this sense provide valuable information for separating the 
different components along the line of sight in the central plane regions. 

\section{Bulge contamination}

\begin{figure}
\begin{center}
\vspace{1cm}
{\par\centering \resizebox*{8cm}{6cm}{\includegraphics{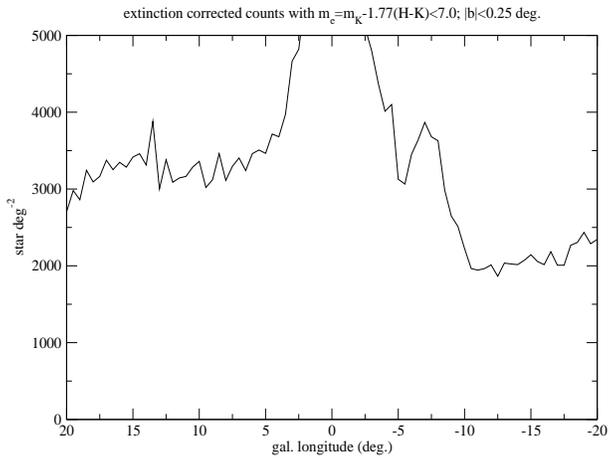}}\par}
\end{center}
\caption{{\itshape 2MASS} star counts corrected for extinction with the data
used in L\'opez-Corredoira et al.\ (2005).}
\label{Fig:m0700b0}
\end{figure}

\begin{figure}
\begin{center}
\vspace{1cm}
{\par\centering \resizebox*{8cm}{6cm}{\includegraphics{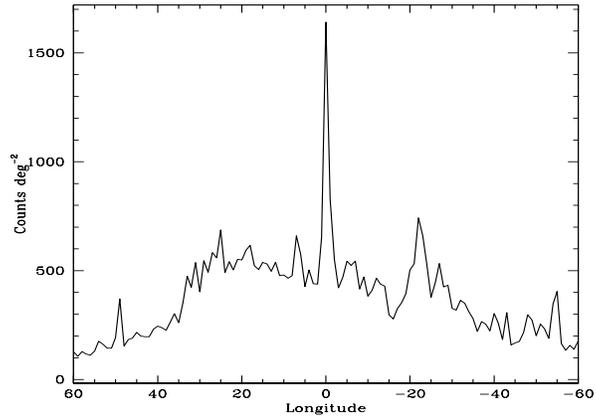}}\par}
\end{center}
\caption{{\itshape MSX} star counts, $m_{[8.7\mu m]}<6.0$, at 
$-0.5^\circ <b<0.5^\circ $, averaged over $\Delta l=1^\circ $,
showing the asymmetry in the plane of positive and
negative longitudes, and the deficit of stars at $l$ between $-15$ and $-18$
degrees (between the hypothetical end at negative longitudes of the bar and the
ring).}
\label{Fig:msx}
\end{figure}

\begin{figure*}
\begin{center}
\vspace{1cm}
{\par\centering \resizebox*{12cm}{8cm}{\includegraphics{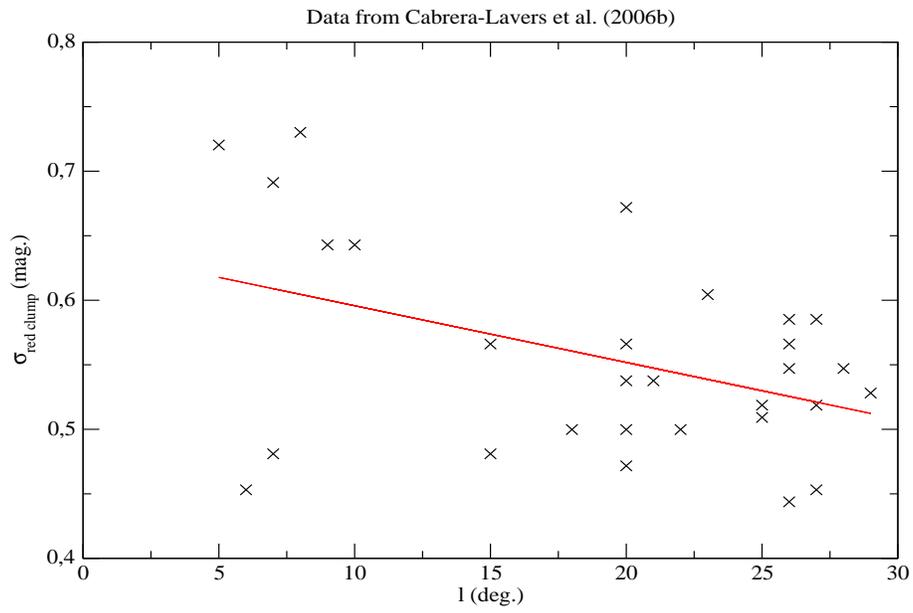}}\par}
\end{center}
\caption{Width of the red clump (Cabrera-Lavers et al., in preparation)
with $|b|\le 1^\circ$. The solid line is the best linear fit: $y=0.64-0.0044\ l$.}
\label{Fig:sigma}
\end{figure*}

\begin{figure*}
\begin{center}
\vspace{1cm}
{\par\centering \resizebox*{12cm}{8cm}{\includegraphics{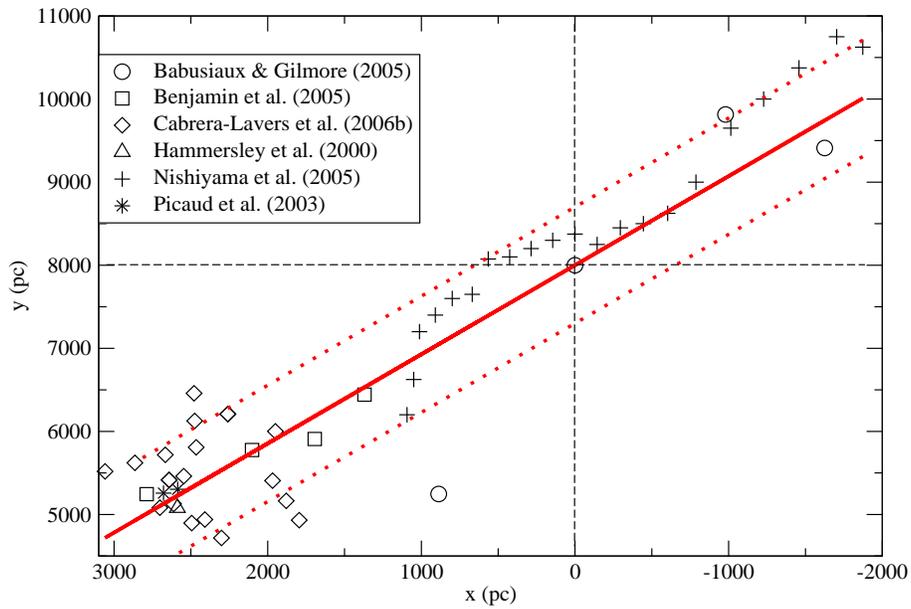}}\par}
\end{center}
\caption{Position of the long bar with respect to the Sun
according to distances derived with the red clump method.
Galactic Centre at $x$=0, $y=8000$ pc.
The H00 bar with a position angle 
of 43$^\circ $ is represented by solid thick lines (and a width
of 1.4 kpc is represented by dotted thick lines).}
\label{Fig:redc}
\end{figure*}

Throughout the present paper we have presented two kinds of data that
show the presence of the bar: 1) counts of bright stars and
2) results from the red clump method. In both cases, 
the presence of the bulge is important only 
for low values of galactic longitude:

\begin{enumerate}

\item The  star counts in Fig.\ \ref{Fig:m0700b0}
give an excess of $\sim$$1000$ stars deg$^{-2}$ in positive
galactic longitudes. According to the models in L\'opez-Corredoira
et al.\ (2005; Model 1), the contribution of the thick bulge in the plane
for these magnitudes is lower than  20\% of this amount for 
$l<-9.5^\circ $ or $l>13^\circ $ and lower than  10\% of this amount for 
$l<-12^\circ $ or $l>17^\circ $. This can indeed be observed in the
Fig.\ \ref{Fig:m0700b0}, where the decreasing density of 
the bulge is not significant away from $|l|$ equal to 10 or 12 degrees.
Therefore, the (nearly flat) excess of counts for  higher  $|l|$ values
than these make no significant contribution to the bulge. 
The same applies
to Fig.\ \ref{Fig:2mass} and the {\itshape MSX} bar counts of Fig.\ \ref{Fig:msx} 
(to be compared with the bulge counts in L\'opez-Corredoira 
et al.\ 2001a).

\item The density of red clump stars is proportional to the total
density of stars. The bulge along the
major axis of the bar has a density $6.6\exp {[-1.1x_1/(740\ {\rm pc})]}$ 
star pc$^{-3}$ (L\'opez-Corredoira et al.\ 2005), where $x_1$ is the distance 
from the center. If
we take the value of the star density in the
bar estimated in \S \ref{.otherp} as a  
$\overline {\rho}=3.5$ star pc$^{-3}$, the bulge would 
be responsible for 20\% of the bar counts at $t=1510$ pc
(equivalent to $l=8.5^\circ $) and 10\% of the bar counts at $t=1970$ pc
(equivalent to $l=11.5^\circ $). For longitudes lower than 8.5$^\circ $
($|x|<1000$ pc in Fig.\ \ref{Fig:redc}) the bulge contaminates more than
  20\% of the bar contribution so we are measuring the position angle of 
a mixture of two populations. 

\end{enumerate}

\section{Conclusions}

Our conclusion is therefore that recent available data on the
long bar corroborate H00's hypothesis.
The Milky Way has a long thin bar with 
approximate dimensions 7.8 kpc $\times $1.2 kpc $\times $0.2 kpc; 
a position angle of the major axis with respect to the line Galactic center-Sun
of approximately 43$^\circ $; density of
stars (brighter in K-band than $M_K=-6.1$) around $7\times 10^{-5}$
pc$^{-3}$, and approximately the third part of stars of the thick bulge
if we assume similar stellar populations for both components.
GLIMPSE data (Benjamin et al.\ 2005) explicitly supported this model,
and other data on the distance of the long bar derived
with the red clump method agree in general with this scenario.

\section*{Acknowledgments}

M. L\'opez-Corredoira was supported by the {\it Ram\'on y
Cajal} Programme of the Spanish Science Ministry. 
This publication makes use of data products from the Two Micron All Sky Survey, 
which is a joint project of the University of Massachusetts and the Infrared Processing 
and Analysis Center/California Institute of Technology, funded by the National Aeronautics 
and Space Administration and the National Science Foundation.
The MSX Point Source Catalog was obtained from the NASA/IPAC Infrared
Science Archive at Pasadena, California.

\appendix

\section{Difference in maximum density along the line of
sight and the position of the major axis}
\label{.dif}

\begin{figure}
\begin{center}
\vspace{1cm}
{\par\centering \resizebox*{3cm}{6cm}{\includegraphics{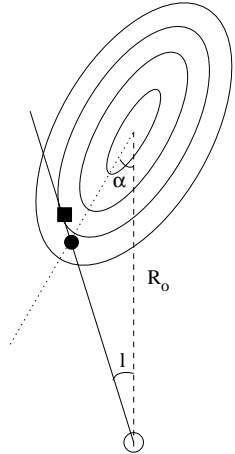}}\par}
\end{center}
\caption{Graphical representation of the difference between the
maximum density position along the line of sight (filled square) for triaxial
ellipsoids and the intersection of their major axes with the line of 
sight (filled circle).}
\label{Fig:eliptan}
\end{figure}

The position of the maximum density 
along the line of sight is not coincident with the 
the position of the major axis of a triaxial structure unless
it be very thin (a bar) rather than a thick bulge. Two
effects produce the difference: i) the most important 
is that the line of sight reaches its maximum at the tangential
point to the innermost ellipsoid and this is not in general
in the major axis (see Fig.\ \ref{Fig:eliptan}); ii) moreover,
in off-plane regions, the lines of sight are not parallel to the
plane $b=0$---the larger the distance the farther away from the plane and the
lower the densities become, so the maximum of the line of sight is indeed
closer than the real maximum in a plane parallel to $b=0$.

We can derive analytically this difference ($\Delta r=r_m-r_a$) between the
maximum density position along the line of sight ($r_m$), and the
intersection of the line of sight with the major axis ($r_a$) for the
case of triaxial ellipsoids with density decreasing monotonically outwards.
The ellipsoidal isodensity contours are defined by the points of space
with the same value of $t$:
\begin{equation}
t=\sqrt{x_1^2+(x_2/A)^2+(x_3/B)^2}
\label{t}
.\end{equation}
$A$ and $B$ are the axial ratios of the second and third axes with
respect to the major axes of the ellipsoids,
and $x_i$ are the cartesian coordinates with the axis of the ellipsoids
centred on the  Galactic Centre; 
that is,
\begin{equation}
x_1=x\sin \alpha -y\cos \alpha,
\end{equation}
\begin{equation}
x_2=x\cos \alpha +y\sin \alpha,
\end{equation}
\begin{equation}
x_3=z .
\end{equation}
We have assumed that the minor axis is perpendicular to the Galactic plane
($x_3=z$). $x$, $y$, $z$ are the cartesian coordinates with $XY$ defining
the plane of the Galaxy, centred on the Galactic Centre, with the
$y$-axis in the Sun--Galactic Centre line. $\alpha $ is the angle between
the major axis of the ellipsoid and this $y$-axis. That is,

\begin{equation}
x=r\sin l \cos b,
\end{equation}
\begin{equation}
y=r\cos l \cos b -R_\odot,
,\end{equation}
\begin{equation}
z=r\sin b
.\end{equation}

The maximum density, the tangential point of the line of sight with
the innermost ellipsoid, i.e.\ maximum density, follows

\begin{equation}
\frac{\partial t}{\partial r}(r_m)=0
\label{der}
.\end{equation}
Equation (\ref{der}) with (\ref{t}) lead together to

\begin{equation}
x_1(r_m)\frac{\partial x_1}{\partial r}(r_m)+
\frac{x_2(r_m)}{A^2}\frac{\partial x_2}{\partial r}(r_m)+
\frac{x_3(r_m)}{B^2}\frac{\partial x_3}{\partial r}(r_m)=0
,\end{equation}
which with all the above expressions lead to
\begin{equation}
r_m=\frac{\frac{R_\odot}{\cos b}\left[\cos \alpha +
\frac{\tan (l+\alpha)\sin \alpha}{A^2}\right]}
{\cos (l+\alpha )+\frac{1}{A^2}\sin (l+\alpha )\tan (l+\alpha )+
\left(\frac{\tan b}{B}\right )^2\frac{1}{\cos (l+\alpha )}}
,\end{equation}
while the position of the major axis (simple geometry with the application
of the sine rule, see Fig. \ref{Fig:eliptan}) is
\begin{equation}
r_a=\frac{R_\odot}{\cos b}\frac{\sin \alpha}{\sin (l+\alpha )}
.\end{equation}

\begin{figure}
\begin{center}
\vspace{1cm}
{\par\centering \resizebox*{6cm}{5cm}{\includegraphics{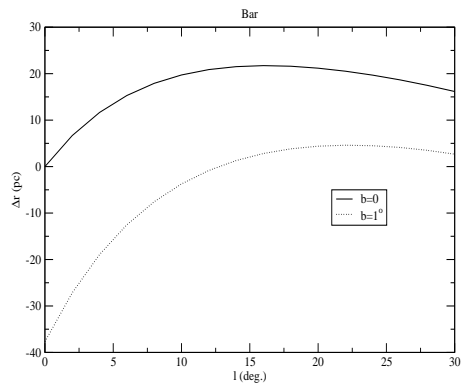}}\par}
\end{center}
\caption{Difference between the
maximum density position along the line of sight for triaxial
ellipsoids of axial ratios 1:0.11:0.04, angle with the line Sun--Galactic
centre 43$^\circ $ (expected for the long bar), and the intersection of 
their major axes with the line of sight.}
\label{Fig:calc_bar}
\end{figure}

Both expressions for $r_m$ and $r_a$ are coincident for $A\ll$ (very 
elongated ellipsoids) and $\tan b\ll$ (in the plane), but the remaining
cases are affected by a non-negligible systematic error, $\Delta r$.
Calculations are carried out for $A=0.11$, $B=0.04$, 
$\alpha =43^\circ $, typical of a long bar:  
Fig.\ \ref{Fig:calc_bar}. The low thickness of the bar (around 1 kpc
thick) is justifiable because of the low dispersion of the red clump
giants (see section \ref{.otherp}). 
The result is that $\Delta r$ is negligible for the bar,
less than $\sim$50 pc of systematic error within $|b|<1^\circ$. 
However, it is not negligible
for the bulge, which reaches a discrepancy of up to 1000 pc
(Cabrera-Lavers et al.\ 2006b).

\section{Difference in maximum density along the line of
sight and the maximum of star counts vs.\ magnitude}
\label{.dif2}

The maximum of star counts vs.\ magnitude is not strictly coincident with the
maximum in density along the line of sight. The distribution
of the magnitude histogram counts in a solid angle $\omega $
is indeed (assuming a constant
absolute magnidude $M$ for all red clump stars and extinction $E(r)$
along the line of sight):

\begin{equation}
A(m)dm=\omega r^2\rho (r)dr
,\end{equation}
\begin{equation}
r=10^{\frac{m-M+5-E(r)}{5}}
\end{equation}
Strictly speaking, we would need to transform $A(m)dm$ into 
$\rho(r)dr$, and then fit its maximum. 
From the relationship between $\rho(r)$ and $A(m)$ it can be deduced
that the difference between the corrected distance to the maximum in 
the density distribution ($r^*_m$ such that $\rho '(r^*_m)=0$) 
and the distance obtained to the maximum in the counts 
histograms ($r_m$ such that $A'(m[r_m])=0$) is [neglecting the term
$E'(r)$]:

\begin{equation}
A(m)\propto r^3\rho (r)
,\end{equation}
\begin{equation}
r^*_m = r_m + \Delta r_m
\end{equation}
\begin{equation}
\Delta r_m = \frac {3 ~\rho (r_m)}{r_m \rho''(r_m)}
.\end{equation}
As $\rho''(r_m)<0$, 
then $r^*_m < r_m $, thus the corrected distances are slightly lower 
than those obtained from the maximum of the magnitude histograms. 
This effect is more noticieable for the bulge component than for the 
long thin bar, as $\rho''(r_m)$ is very large for the bar. Cabrera-Lavers
et al.\ (2006b) have 
estimated the range of values for $\Delta r_m$ both for the bulge and 
for the bar, and have found the true density distributions along the line of sight. 
$\Delta r_m$ ranges from 100 to 300 pc in bulge fields, whereas for bar fields
the effect is almost negligible
with corrections in the order of 25--50 pc.

\eject

\end{document}